# Interleaved Electro-Optic Dual Comb Generation to Expand Bandwidth and Scan Rate for Molecular Spectroscopy and Dynamics Studies near 1.6 μm

JASPER R. STROUD[1], JAMES B. SIMON[2], GERD A. WAGNER[3], AND DAVID F. PLUSQUELLIC[1]

[1] *Applied Physics Division, Physical Measurement Laboratory, National Institute of Standards and Technology, Boulder, CO 80301.*
[2] *Department of Physics, University of California, Berkeley, CA 94720.*
[3] *Deutsches Zentrum für Luft- und Raumfahrt (DLR), Institute of Technical Physics, Pfaffenwaldring 38-40, 70569 Stuttgart, Germany*
*david.plusquellic@nist.gov*

**A chirped-pulse interleaving method is reported for generation of dual optical frequency combs based on electro-optic phase modulators (EOM) in a free-running all-fiber based system. Methods are discussed to easily modify the linear chirp rate and comb resolution by more than three orders of magnitude and to significantly increase the spectral bandwidth coverage. The agility of the technique is shown to both capture complex line shapes and to magnify rapid passage effects in spectroscopic and molecular dynamics studies of $CO_2$. These methods are well-suited for applications in the areas of remote sensing, reaction dynamics, and sub-Doppler studies across the wide spectral regions accessible to EOMs.**

## 1. Introduction

Dual optical frequency comb (DOFC) systems based on mode-locked lasers (MLL) [1,2] are now well-established and proven to provide significant advantages in numerous applications across the spectrum from the acoustic to the deep-UV regions [3,4,5,6,7,8]. Although typically operated at significantly reduced spectral coverage, electro-optic dual optical frequency combs (EO-DOFC) are an attractive alternative owing to the simplicity of the methods to generate the dual combs from a single free-running laser while maintaining mutual phase coherence in the down-converted radiofrequency (RF) region [9,10,11,12,13,14,15,16]. In many areas, the limited bandwidth of a few $cm^{-1}$ is advantageous to enhance sensitivity in photon counting applications, to significantly increase comb resolution for quantum state resolved studies of single absorption bands [17,18,19,20], or for the sub-Doppler coherent interrogation of atoms [21,22].

EO-DOFCs provide agility in a few key areas. The comb tooth spacing (i.e., repetition rate) can easily extend over orders of magnitude by simple changes to the electronic waveform generator driving the EO modulators (EOMs) [23]. While broadband self-referenced EO combs have been generated that cover thousands of wavenumbers with 10s of GHz comb spacings [24,25,26], some applications benefit from the resolution agility over a few wavenumbers. For example, standoff methods for remote sensing of greenhouse gases (GHGs) in the atmosphere including differential absorption LIDAR (DIAL) [27,28] and natural target integrated path differential absorption LIDAR (IPDA) [29,30] typically require photomultipliers or avalanche photodiodes [31,32] to detect faint backscattered returns and therefore, require high power per comb tooth over bandwidths of (2-5) $cm^{-1}$ and at comb resolutions from 100 MHz to 250 MHz [33]. Single quantum level studies of molecules prepared in cold environments also require (2-5) $cm^{-1}$ of spectral coverage to investigate a single rovibrational [34] or rovibronic band [35,36,37] but with sub-MHz resolution. Under these conditions, achieving the highest signal-to-noise ratio for a given source power and measurement time requires limiting the bandwidth coverage.

The bandwidth expansion of EO-DOFCs to several wavenumbers (100's of GHz) from microwave (MW) sources operating below 1 $cm^{-1}$ (30 GHz) has recently been demonstrated using different interleaving schemes. A few methods have made use of two cascaded EOM stages [38] or injection locking schemes [39] to superimpose fine comb lines on a widely spaced LO comb. Spectral coverage up to 360 GHz [39] and 450 GHz [40] have recently been reported using these methods. Other techniques have made use of high-power pulsed sources (such as erbium doped fiber amplifiers, EDFAs) to expand coverage over (10-100) $cm^{-1}$ [20,23].

An alternative approach to extend coverage is to drive the EOMs with chirped pulse waveforms from an arbitrary waveform generator (AWG). Although flattened EOM combs can be obtained using a number of methods including cascaded intensity and phase modulators [41] or dual-drive Mach-Zehnder modulators [42], the use of linear frequency chirps can provide an even (and easily controllable) distribution of the source power over the chirp bandwidth. The downside of chirped pulse approach is the inability to resolve the orders in the optical spectrum since all optical orders overlap when mapped to the RF region in the down-conversion process [22]. For single frequency drives, EOMs generate (+) and (-) sidebands at frequency, $\Omega$, on the carrier field, $E_0\exp(i\omega t)$, and have sideband amplitudes for the different orders, $k\Omega$, given by

$$E_0 e^{i\left[\omega t+\beta\sin(\Omega t)\right]} = E_0 e^{i\omega t}\left[ J_0(\beta) + \sum_{k=1}^{\infty} J_k(\beta) e^{ik\Omega t} + \sum_{k=1}^{\infty} (-1)^k J_k(\beta) e^{-ik\Omega t} \right] \qquad (1)$$

where $J_k(\beta)$ is the Bessel function of order $k$, and $\beta$ is the peak MW voltage relative to the modulator's $V_\pi$ ($V_\pi$ is the half-wave voltage required to cause a π phase shift). The expanded coverage is given by $\pm k\Omega$, where $\Omega$ is the microwave (MW) frequency driving the EOM. While the (+) and (-) sidebands are easily separable using an AOM on one or both comb legs of the interferometer, the orders of the EOM are more difficult to separate. Typically, to reduce contamination from the higher order modes, low MW amplitudes ($<V_\pi/4$) are used which limits the $1^{st}$-order coverage of EOMs to twice the bandwidth of the MW source [43,44]. To take advantage of the order scaling of chirped pulse EO-DOFCs, a new method for interleaving [45,46] is needed to ensure the unique one-to-one mapping of each order into the RF region.

An additional element of agility of chirped pulse EO-DOFCs is the precise control of the speed of coverage or sweep rate which can be easily changed by several orders of magnitude depending on the desired application [23]. Direct applications of chirped pulse sources in the MW region [47] and with extension into the THz using amplifier/multiplier chains [48,49] are now routinely used in gas phase studies for coherent quantum state preparation and deep signal averaging of the time-domain response over a few hundred GHz. Furthermore, chirp times from a few µs down to a few tens of ns enable sensitive background-free detection of the free induction decays. Similar to that demonstrated in the mid-IR using cw [50] and pulsed quantum cascade lasers (QCL) [51,52,53,54], here we show that AWG driven EO-DOFCs can reach the chirp rates necessary to extend these MW/THz techniques to broad bandwidths in the near-IR [55,56,21] and beyond.

This paper is structured as follows. In Section 2 (experimental setup) and Section 3 (methodology), two methods are presented and discussed to broaden the spectral coverage by effectively interleaving the orders of the (+) and (-) sidebands. One method is based on a segmented scan approach [48] to cover the $\pm k\Omega$ scan range while limiting the IF bandwidth for down-conversion to less than 500 MHz. The second method is based on a dual chirp scheme which can be used to significantly increase the optical scan rate at the expense of comb resolution. Results of the two methods are provided in Section 4. In Section 4.1, first order spectra and analysis are shown for an optical bandwidth coverage of 30 GHz (1 $cm^{-1}$), and Section 4.2 shows higher-order spectra (up to $4^{th}$-order) with an optical bandwidth coverage of up to 120 GHz (4 $cm^{-1}$) to sample simultaneously four sequential $CO_2$ lines at 1.6 um with scan times as short as 8 ms. Section 4.3 provides signal-to-noise ratio (*SNR*) estimates and discusses further improvements. Section 4.4 demonstrates increased scan rates as high as 43.5 MHz/ns and discusses in detail the effects of rapid passage on the observed line shapes using a Maxwell-Bloch modeling approach. The paper closes with a summary and an outlook in Section 5.

## 2. Experimental setup

The electro-optic dual comb system is shown in Fig. 1. An external cavity diode laser (ECDL, New Focus, Model 6330) [57] is used as the seed source and is actively locked using a Pound-Drever-Hall (PDH) feedback to an external 0.5 m confocal reference cavity [58] (Burleigh, finesse~100, FSR=300 MHz, Invar spacer) to achieve frequency stability of < ±100 kHz. The laser output is then split into two legs, and feeds two fiber coupled acousto-optic modulators (AOMs, Brimrose) operating near 50 MHz. The two AOMs are driven at slightly different frequencies with two function generators (Keithley, 3390) with differences up to 1 MHz. The output of the AOMs are then coupled into two EOMs (EO-space, PM-5K4-10-PFA-PFA-UV-UL), each driven by a channel of the AWG (Keysight, M8195). The two AWG channels are programmed to define a local oscillator (LO leg) and signal (SIG leg) waveforms. The SIG leg is free-space coupled and double-passed through a one-meter evacuable sample cell and then combined with the LO leg output. The combined output is detected by a 400 MHz balanced avalanche photodiode (Thorlabs, PDB570C). The output signal is split using a two-way power splitter (ZSC-2-1+), where one side is digitized at 1 GS/sec with 12-bit digitizer card and the other side sent to a phase locking circuit.

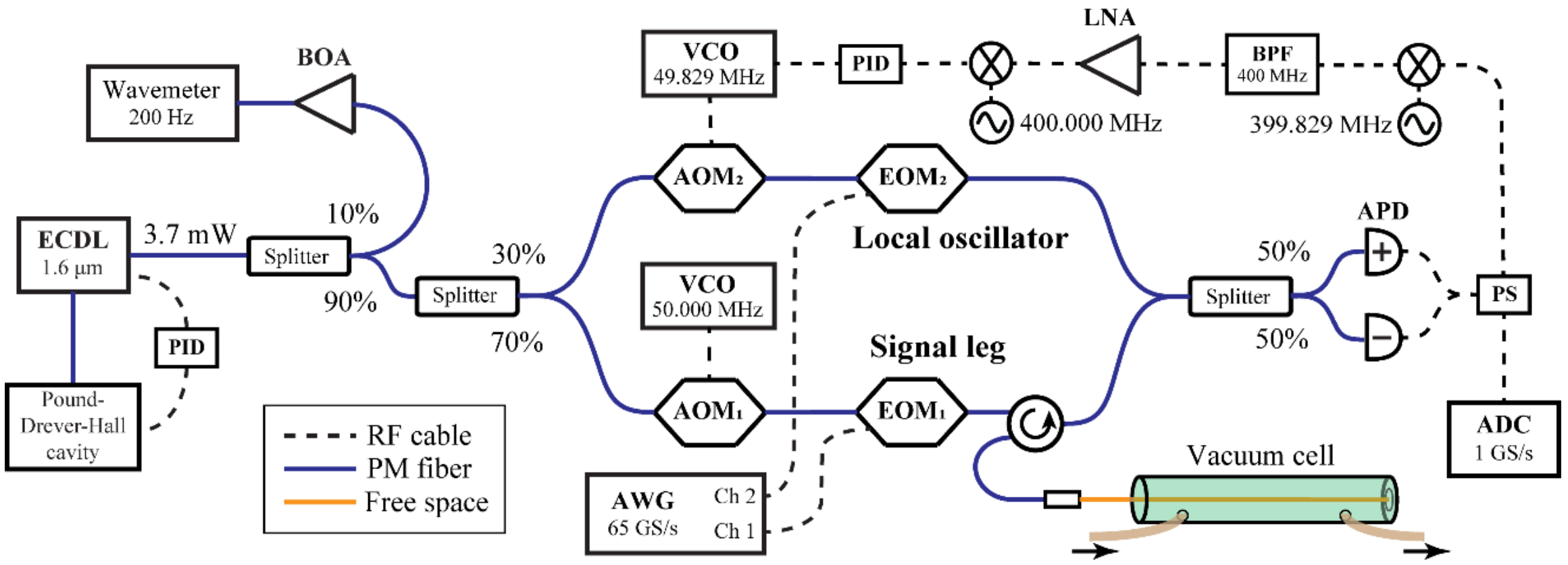

**Figure 1**. Schematic of the interleaved EO-DOFC spectrometer. A small component (5 %) of the external cavity diode laser (ECDL) output is used to frequency stabilize the laser to < ± 100 kHz. A second small pick (10 %) is split with part sent to a wavemeter (10 %) and part (90 %) split again in a 30 %/70 % fiber splitter to define the local oscillator (LO) and signal legs, respectively. Each leg is coupled to an AOM followed by an EOM. The signal leg is free-space coupled and double-passed through a one-meter vacuum cell and combined with the LO leg and detected by a 400 MHz balanced avalanche photodiode (APD). The output of the APD is power split with part sent to a digitizer and part mixed with 399.829 MHz tone. The upconverted signal is filtered, amplified by 20 dB and mixed with a 400 MHz tone. The error signal is conditioned in a proportional-integral-derivative circuit that drives a voltage-controlled oscillators (VCO). The VCO drives $AOM_2$ to maintain phase lock of the beat frequency, $\Delta f_{AOM}$, at the detector. See text for details.

The difference in the AOM driving frequencies (171 kHz shown in Fig. 1) defines a beat note that serves to interleave the EOMs (+) and (-) sidebands (*vide infra*). For long term stability, the beat note needs to be stabilized. To do this, the beat note is mixed with a 399.829 MHz LO (local oscillator, SRS, SG-382, phase noise < 120 dbc $Hz^{-1}$) and passed through a 6-pole electronic filter (K & L Microwave, 6C30-400/T0.8-0/0, 850 kHz 3 dB point) centered at 400 MHz. The filter output is amplified by +17 dB (Mini-circuits, ZFL1000+) and sent to a high frequency mixer (Holzworth, HX3100) referenced to a 400.000 MHz LO (SRS, SG-382) to generate an error signal. The error signal is first low passed at 10 kHz and conditioned in a PID servo module (Vescent Photonics, D2-125). The servo output drives the frequency modulation input to one of the function generators driving the AOMs to phase correct the signal at the balanced detector. All MW components are phase locked to a 10 MHz Rb reference.

The acquisition system consists of a dual channel 12-bit, 3-GS/s acquisition card (Gage, CSE123G2). The throughput is optimized to acquire data on two channels at 1 GS/sec (streaming rate of 4 GB/s) and processed in real time using a computer system with 56 logical processors providing a bandwidth for the acquisition of RF signals up to 500 MHz. Circular memory buffers are used to optimize throughput and for this work, 50 buffers are used to provide 400 MS of total acquisition memory (12-bit sample size). The size and number of buffers are important when choosing AOM frequencies and determining the trigger rate as discussed below.

## 3. Methodology

Fig. 2 shows the general concept of the interleaving method. The signal leg is chirped over the frequency range, $f_{start}$ to $f_{stop}$, in *j* segments to sample the spectral region around the diode laser frequency (ECDL). To map this near-IR region onto the detectable RF frequency, the LO asynchronously follows the SIG leg chirp. For each segment of the SIG leg chirp, the LO is chirped from $f_{LO_{start_j}} = f_{start_j} - f_{IF_{start}}$ to $f_{LO_{stop_j}} = f_{stop_j} - f_{IF_{stop}}$, generating beat notes on the detector that map the IR spectrum from $f_{start_j}$ to $f_{start_j}$ to the RF frequencies $f_{IF_{start_j}}$ to $f_{IF_{stop_j}}$. Furthermore, each chirp segment is repeated a total of

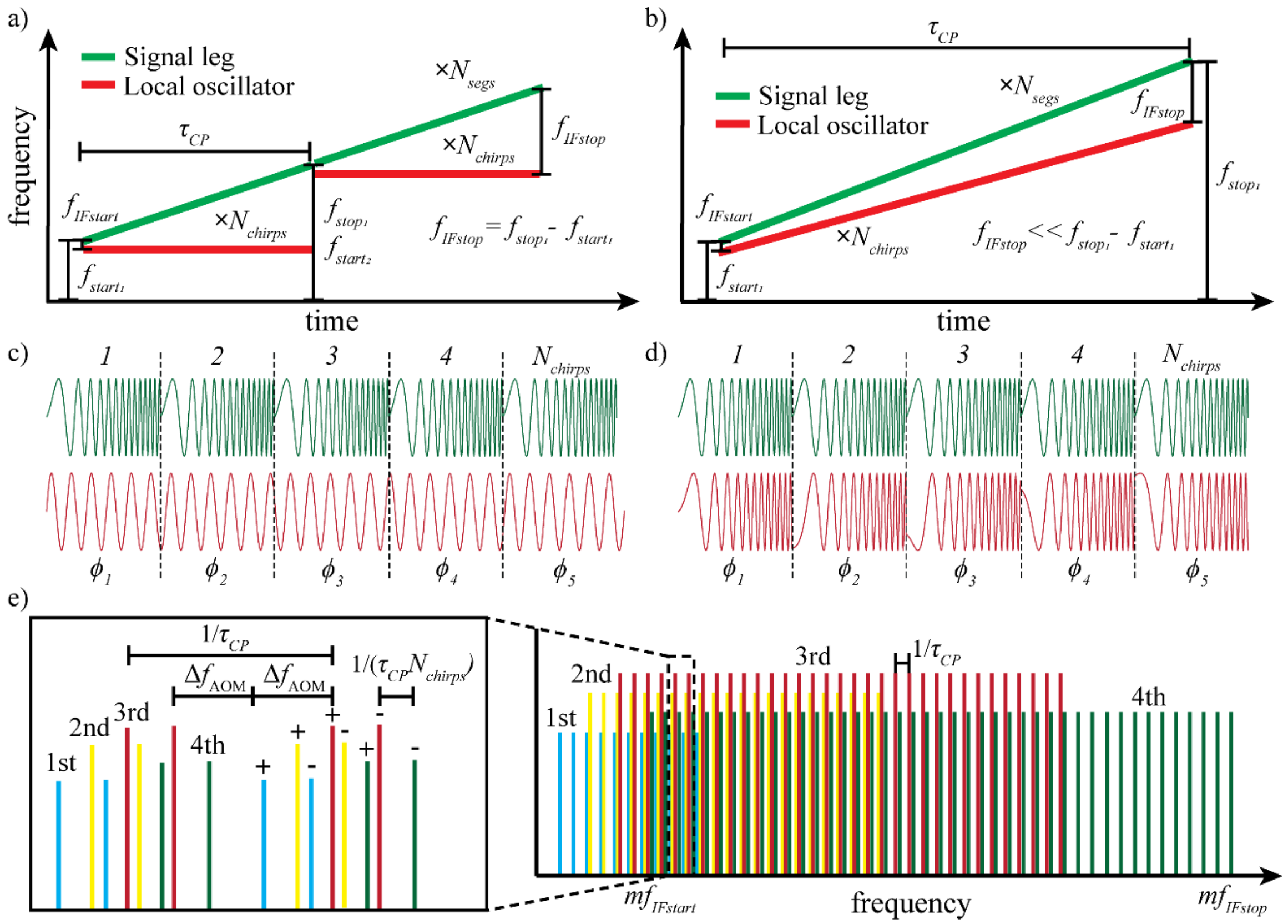

**Figure 2**. Conceptual illustration of the two methods using a) a segmented scan approach (two segments shown) having a fixed LO frequency that results in the RF comb resolution that equals the optical resolution, and b) the dual chirped technique where the optical resolution is much lower than the RF comb resolution (only one segment shown). The phase slip on the LO leg is shown in c) for a fixed LO and d) for a chirped LO where the dashed lines indicate consecutive shifts that sample across the full beat note. e) The location of the comb teeth is shown in relation to the chirp time $\tau_{CP}$, number of chirps, $N_{chirps}$, and the AOM beat note, $\Delta f_{AOM}$.

$N_{chirps}$ with a different phase shift applied to the LO for each chirp cycle, $i$. The waveforms on the LO and SIG legs can be generally described as,

$$WF_{i,j}^{SIG}(t) = \sin\left(2\pi f_{start_j} t + \frac{2\pi(f_{stop_j} - f_{start_j})}{2\tau_{CP}} t^2\right) \tag{2}$$

$$WF_{i,j}^{LO}(t) = \sin\left(2\pi f_{LO_{start_j}} t + \frac{2\pi(f_{LO_{stop_j}} - f_{LO_{start_j}})}{2\tau_{CP}} t^2 - \frac{2\pi LO_{mult}}{N_{chirps}}\left(\frac{t}{\tau_{CP}} + i\right)\right) \tag{3}$$

$$i = 1 \ldots N_{chirps}$$

$$j = 1 \ldots N_{segs}$$

$$t = 0 \ldots \tau_{CP}$$

where $\tau_{CP}$ is the duration for a single chirp, and $LO_{mult} = 1, 2, 3\ldots$ is an integer multiplier corresponding to the frequency shift separating the different comb orders. The RF frequency comb is generated by Fourier transforming a repeated series of chirps over $N_{chirps}$ and $N_{bufs}$. The IF resolution (i.e., comb teeth spacing) is defined by the inverse of the repetition rate in the time domain,

$$\Delta f_{IF_{res}} = \tau_{CP}^{-1} \tag{4}$$

while the optical resolution is defined as

$$\Delta f_{Opt_{res}} = \left[1 - \left( f_{LO_{stop_j}} - f_{LO_{start_j}} \right) \Big/ \left( f_{stop_j} - f_{start_j} \right)\right]^{-1} \Delta f_{IF_{res}} \tag{5}$$

We also define three relationships to help improve clarity in the discussion.

$$\Delta f_{chirp} = \frac{\left( f_{stop} - f_{start} \right)}{N_{segs}} \tag{6}$$

$$\tau_{buffer} = N_{chirps} N_{segs} N_{rep} \tau_{CP} \tag{7}$$

$$\tau_{WF} = \tau_{buffer} / N_{rep} \tag{8}$$

where $N_{rep}$ is used to increase the chirp rate while maintaining a fixed buffer size for data acquisition. The buffer time is fixed at $\tau_{buffer}$ = 8 ms to maintain constant record lengths over the different data sets. If the time of the waveform record, $\tau_{WF}$, is less than $\tau_{buffer}$, the waveform is repeated by $N_{rep}$ to fill the buffer. Figure 2a and 2b illustrate how the SIG and LO waveforms are related in time vs. frequency to down-convert the optical spectrum scanned by the SIG leg. For small scan ranges when the chirp terms in the signal leg are $f_{stop_j} - f_{start_j} = f_{IF_{stop}} - f_{IF_{start}}$, the LO is a fixed frequency and the chirp term disappears in Eq. 3 since $f_{LO_{start_j}} = f_{LO_{stop_j}}$. This also results in equal RF and optical frequency resolutions from Eq. 5. Chirping over larger ranges than the RF detector bandwidth requires chirping the LO asynchronously relative to the SIG leg, which results in a decrease in the optical frequency resolution relative to that of the RF comb according to Eq. 5.

In order to separate the positive and negative sidebands generated by the EOMs, the SIG and LO legs are each shifted by AOMs at different frequencies to generate a beat note in the kHz range. To separate the comb teeth in each of the orders of the EOM, the SIG leg is scanned multiple times, $i = 1 \ldots N_{chirps}$, while a linear phase shift is applied to the LO leg, $-\left(t/\tau_{CP} + i\right) 2\pi LO_{mult} / N_{chirps}$ where $LO_{mult}$ is an integer scalar for the phase shift. Figure 2c and 2d shows how this linear phase shift changes the relationship between the LO and the SIG waveforms as the SIG chirp is repeated for a fixed and chirped LO, respectively. Over the $N_{chirps}$, this phase slip generates a corresponding frequency shift that scales with EOM order, shown as the final term in Eq. 9. The location of the comb teeth is determined by,

$$f_{comb}(n,k) = \pm \Delta f_{AOM} + n \Delta f_{IF_{res}} + k f_{IF_{start}} + \frac{k LO_{mult} \Delta f_{IF_{res}}}{N_{chirps}} \tag{9}$$

$n$ labels the comb tooth, $k$ is the order and $\Delta f_{AOM}$ is the beat note defined by the AOMs. The $\Delta f_{AOM}$ beat note should not share a common denominator with $\Delta f_{IF_{res}}$ to avoid order overlap. This interplay is illustrated in Fig. 2e, where the location of each comb line should not overlap with the others to successfully demodulate the different comb orders.

The acquisition and concatenation of 50 consecutive buffers increases the digital resolution of the resulting Fourier transform. The AWG and the data acquisition system are synchronized using external triggers and require a small reset period for rearming between buffers. Furthermore, each of the buffers needs to be interleaved such the recovered beat note, $\Delta f_{AOM}$ is continuous and fully sampled. This condition requires a relationship between the beat note and the total phase shifted chirp time, $\tau_{CP} N_{chirps}$, such that a fractional phase shift occurs between buffers. To enforce this condition, the trigger rate for the AWG and data acquisition systems is defined as:

$$f_{trig} = \left( \tau_{buffer} + \frac{C_{slip}}{\Delta f_{AOM}} \right)^{-1} \qquad (10)$$

where the total record time $\tau_{buffer} N_{segs}$ is a multiple of the $\Delta f_{AOM}$ beat note and the phase slip introduced by the term $C_{slip} = k_1 + x$ is related to the total phase-shifted chirp time by,

$$\Delta f_{AOM} \tau_{CP} N_{chirps} = k_2 + x \qquad (11)$$

where $k_1$ and $k_2$ are any integers and $x$ is a fractional phase slip. Furthermore, the specific section of the beat note sampled must remain constant for each of the 50 buffers to coherently add in the time domain, requiring $xN_{bufs}$ to also be an integer. We also note that other sampling schemes can introduce this phase slip, for example, by running the AWG in continuous mode and adjusting the buffer size at the required trigger rate.

## 4. Results

The interleaved dual comb system was used to scan over $CO_2$ lines around 1.6 um in a 1.12 m double passed cell ($L_{path}$ = 2.2 m) at variable pressure. Segmented chirps extended over the full 15 GHz bandwidth of the dual channel AWG. The spectral coverage of the **(±)** sidebands centered at the laser frequency was 30 GHz, 60 GHz, 90 GHz, and 120 GHz (4 cm$^{-1}$) for the 1$^{st}$-, 2$^{nd}$, 3$^{rd}$, and 4$^{th}$-orders, respectively.

### 4.1 First order spectra

A segmented scan method was first performed to cover a region from $f_{start}$ = 500 MHz to $f_{stop}$ = 10 GHz. The parameters for the scan include $\tau_{CP}$ = 20 μs, $N_{chirps}$ = 10, and $N_{segs}$ = 40 that define a buffer duration of $\tau_{buffer}$ = 8 ms. The phase slip parameters of $C_{slip}$ = 1.2 cycles and $\Delta f_{AOM}$ = 171 kHz were used to give $f_{trig}$ = 128.890447 Hz and a frequency step size of $\Delta f_{chirp}$ = 237.5 MHz. In this case, the RF comb resolution matches the optical resolution, $\Delta f_{IF_{res}} = \Delta f_{Opt_{res}}$ = 50 kHz (see Tables in the Supplemental Materials for parameter summaries). The FFT of each segment after sorting the buffers resulted in 10 MS records and a transform limit of 100 Hz. The widths of the comb lines are narrower than this transform limit.

The spectra obtained for $CO_2$ at 6241.4770 cm$^{-1}$ are shown in Figs. 3a-3c. The total average power of the EO-DOFC was limited to 200 μW (an average of ≈ 0.5 nW per comb line) to prevent saturation at the balanced detector output. The spectra were acquired in ≈ 40 sec which corresponds to 100 coherent additions of the circular buffers and a total acquisition of 40 GS. The $CO_2$ gas spectra was measured at 6.6 kPa (50 Torr) and at room temperature (294 K). A complex Voigt function based on the Faddeeva function was used to model the spectrum [59,60] (details of the line shape models are given in the Supplemental Materials). Superimposed on the experimental data in Figs. 3a-3c are the best fit spectra for the real and imaginary parts of the Voigt profile.

Given the overall shape of the residuals, the quadratic speed-dependent Voigt profile (qSDVP) [61] was used to improve the fit to the data in Fig. 3. This line shape function requires two additional speed-dependent shift and broadening parameters. Details on the line shapes are included in the supplementary document. The Voigt profile was defined using the measured temperature, pressure and path length while the SDVP profile was fit using a nonlinear least-squares Levenberg-Marquardt algorithm to determine the speed dependent line shape parameters. The fits using the qSDVP line shape are shown in Fig. 3b and 3c, resulting in a slight reduction of the residuals over the $CO_2$ absorption line. We note that the finer structure in the residuals likely have contributions from standing wave interference in the free-spaced cell, phase instabilities over the 40 sec sampling interval, and dynamic range limitations associated with the detector and digitizer.

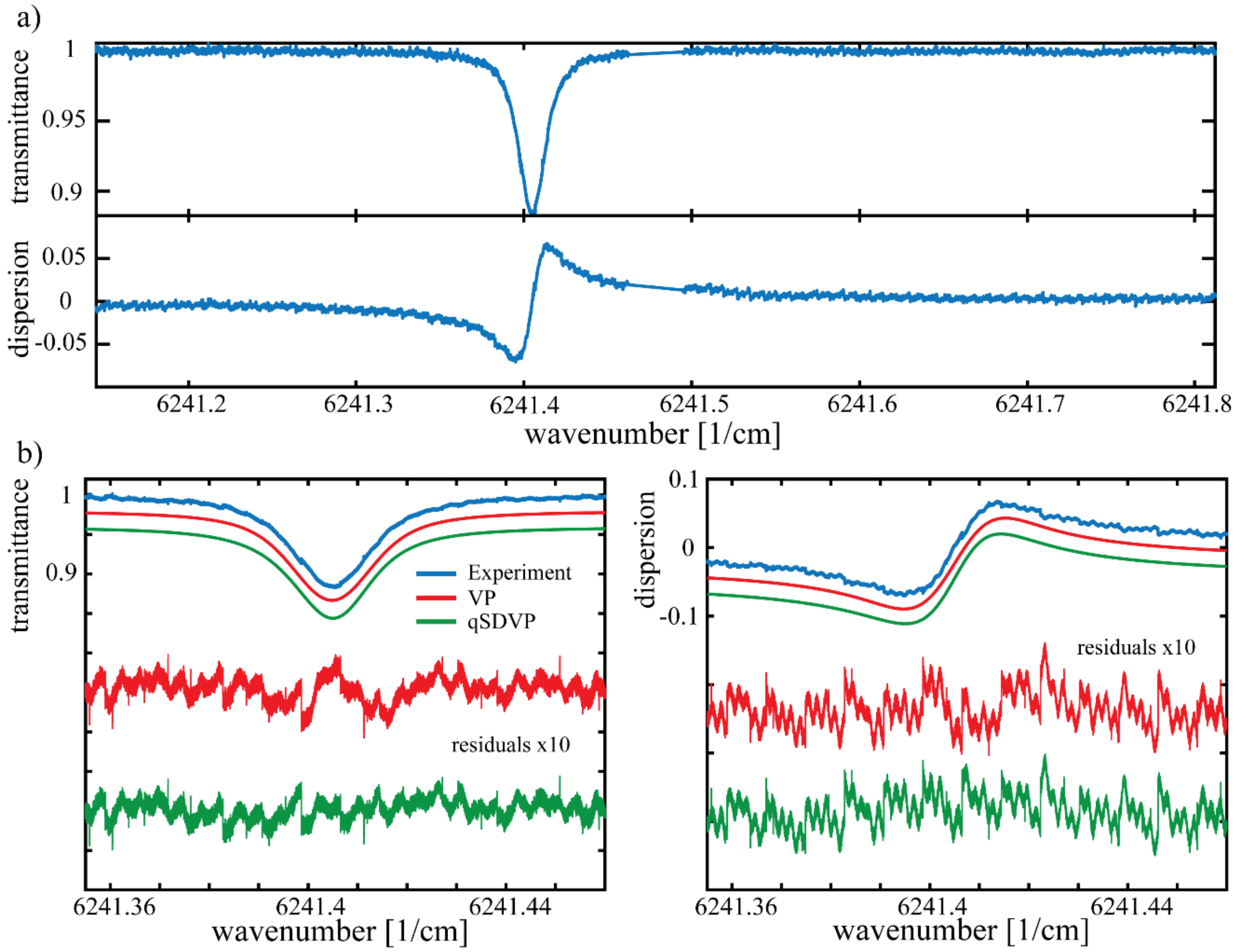

**Figure 3**. First order results showing a) the amplitude and phase spectrum of the $CO_2$ line at 6241.4 cm$^{-1}$ (30013←00001 R 18e) b) The best fits using the Voigt profile (VP) in red and the quadratic speed-dependent Voigt profile (qSDVP) in green, showing slight improvement to a standard VP. The fits and residuals (x10) are offset for clarity.

### 4.2 High order spectra

To generate higher order sidebands on the EOMs, the RF drive power was increased to ≈ $3V_{\pi}$ and the driving amplitudes across the chirp were adjusted to level the output of the 4$^{th}$-order comb lines. Every order has the same spectral comb resolution, but the regions covered by the IF and optical spectra scale with order. To properly allocate the required bandwidth for the 4$^{th}$-order comb, we choose $f_{IF_{start}}$ = 5 MHz, and $f_{IF_{stop}}$ = 100 MHz, so the 4$^{th}$-order IF spectrum will span 20 MHz to 400 MHz. The spectra shown in Fig. 4a are scanned from $f_{start}$ = 500 MHz to $f_{stop}$ = 15 GHz with $N_{segs}$ = 40. The frequency step size was $\Delta f_{chirp}$ = 362.5 MHz and $\tau_{CP}$ = 20 μs to again give $\Delta f_{IF_{res}}$ = 50 kHz. The 3$^{rd}$ and 4$^{th}$-order spectra in Fig. 4a were acquired in segmented mode at room temperature and at a pressure of 6.8 kPa (51 torr). With the laser centered at 6241.9928 cm$^{-1}$, the 4$^{th}$-order spectrum covers four $CO_2$ absorption lines with $\Delta f_{Opt_{res}}$ ≈ 191 kHz and > 600,000 comb teeth. The residuals from the fits are also shown below the spectra.

This same region was acquired in Fig. 4b at a 4-fold faster rate using with $N_{segs}$ = 10, and $LO_{mult}$ = 4. This also results in a 4-fold reduction in resolution ($\Delta f_{Opt_{res}}$ ≈ 763 kHz) compared with Fig. 4a. Both spectra were acquired with multiple segments and chirped over the period, $\tau_{CP}$ = 20 μs, resulting in the same 50 kHz comb at the detector.

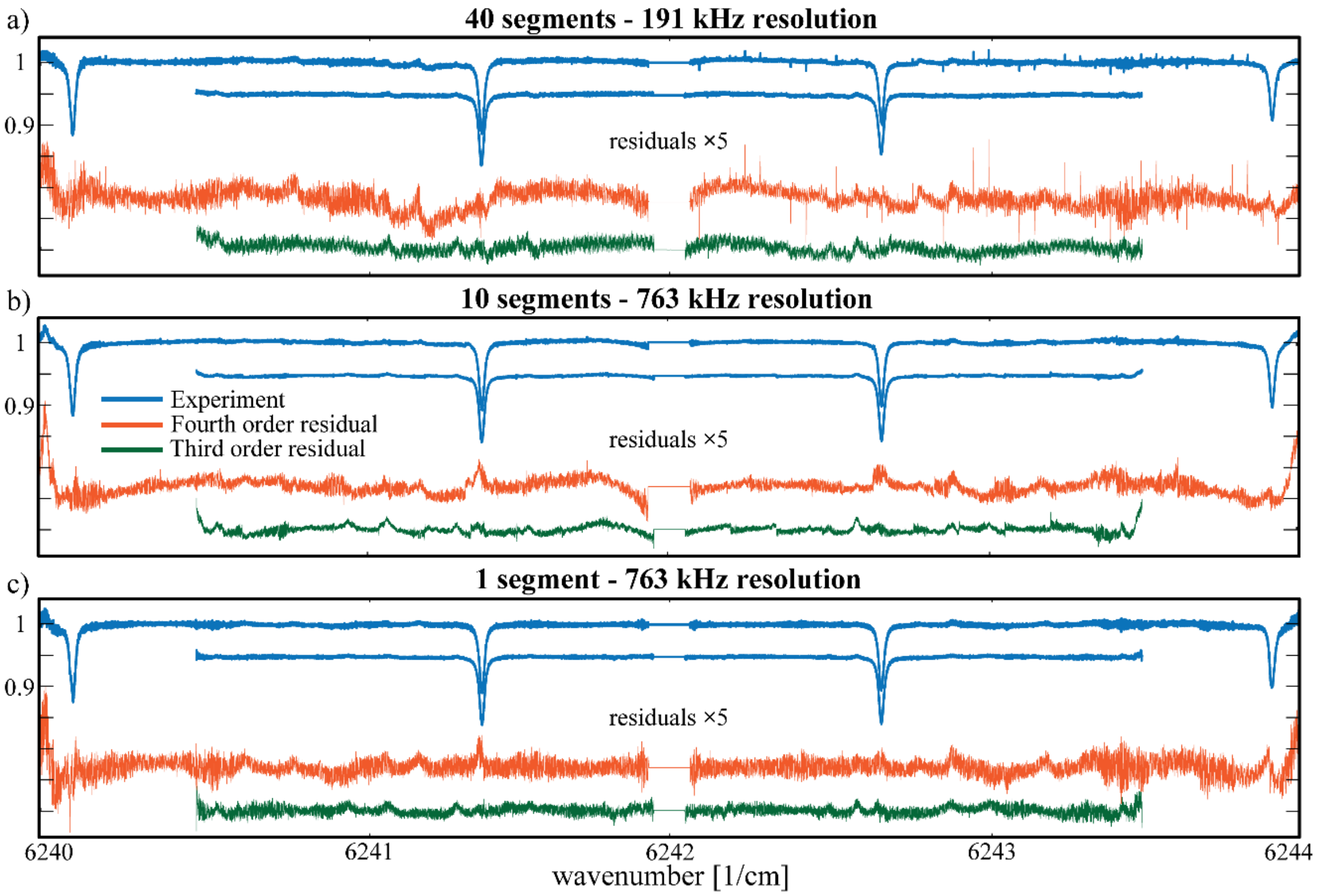

**Figure 4**. The 3$^{rd}$- and 4$^{th}$-order spectra of $CO_2$ (30013←00001, R16e, R18e, R20e, R22e) using a) $N_{segs}$ = 40 segments, $LO_{mult}$ = 1, and $\tau_{CP}$ = 20 μs to give 191 kHz resolution, b) $N_{segs}$ = 10 and $LO_{mult}$ = 4 to give 763 kHz resolution and c) one segments, $LO_{mult}$ = 4, and $\tau_{CP}$ = 200 μs to give 763 kHz resolution. The qSDVP fits and residuals (offset for clarity) are shown in red and green for the 4$^{th-}$ and 3$^{rd}$-order fits, respectively.

In Fig. 4c, the entire spectral range is covered in a single scan ($N_{segs}$ = 1). In this case, we choose $LO_{mult}$ = 4, $N_{chirps}$ = 40, and $\tau_{CP}$ = 200 μs. The AOM beat note is adjusted to $\Delta f_{AOM}$ = 171.2525 kHz, and all 50 buffers are used for a single FFT having a transform limit of 2.5 Hz. These conditions allow for the AOM beat note to be fully sampled. The resolution of the RF combs is $\Delta f_{IF_{res}}$ = 5 kHz, and the different orders are separated by 500 Hz. The 3$^{rd}$- and 4$^{th}$-order spectra and fits are shown in Fig. 4c at the same optical resolution as in Fig 4b ($\Delta f_{Opt_{res}}$ ≈ 763 kHz), since the number of segments, $N_{segs}$ and chirp time $\tau_{CP}$ are commensurately decreased and increased, respectively.

### 4.3 Signal-to-Noise estimates

With interleaving, all orders are acquired simultaneously. However, since high-order generation in the EOMs is MW power dependent, the *SNRs* of different orders depend on the relative powers at the detector and are determined by Bessel function amplitudes for those orders as given in Eq. 1. The optical powers in each of the two legs were approximately equal when using the 30%/70% beam splitter (see Fig. 1). The noise equivalent absorption is estimated using [20]

$$NEA = (L_{abs} SNR)^{-1} \left( \frac{T}{M} \right)^{1/2} \qquad (12)$$

where the $L_{abs}$ is the pathlength, $M$ is the number of comb lines and $SNR$ is the signal-to-noise ratio over the averaging time, $T$. Since these spectra were obtained simultaneously, the total number of comb lines, $M$, in the 3$^{rd}$- and 4$^{th}$-orders is ≈ 275,000. From the residuals shown in Fig. 4b, the *SNR*s of the 3$^{rd}$- and 4$^{th}$-order modes are estimated at 580 and 350, respectively. For $T$ = 40 sec and $L_{abs}$ = 220 cm, the *NEA*s for these two orders are $9.4\times10^{-8}$ cm$^{-1}$/Hz$^{-1/2}$ and $1.6\times10^{-7}$ cm$^{-1}$/Hz$^{-1/2}$, respectively. It is noted that the *NEA* estimates do not account for the weaker comb lines present from lower and higher order modes. Current efforts are underway to improve the *SNR* by using the in-phase and quadrature inputs of dual parallel Mach-Zehnder EOMs to enhance the output in specific orders. We also note that for averaging times of more than a few seconds, the *SNR* is reduced slightly from variations in the mutual phase coherence between the two legs. Actively stabilizing the sample cell length, locking the laser to a self-referenced comb source, and including a phase correction algorithm in the on-line processing code are also expected to improve the *SNR*.

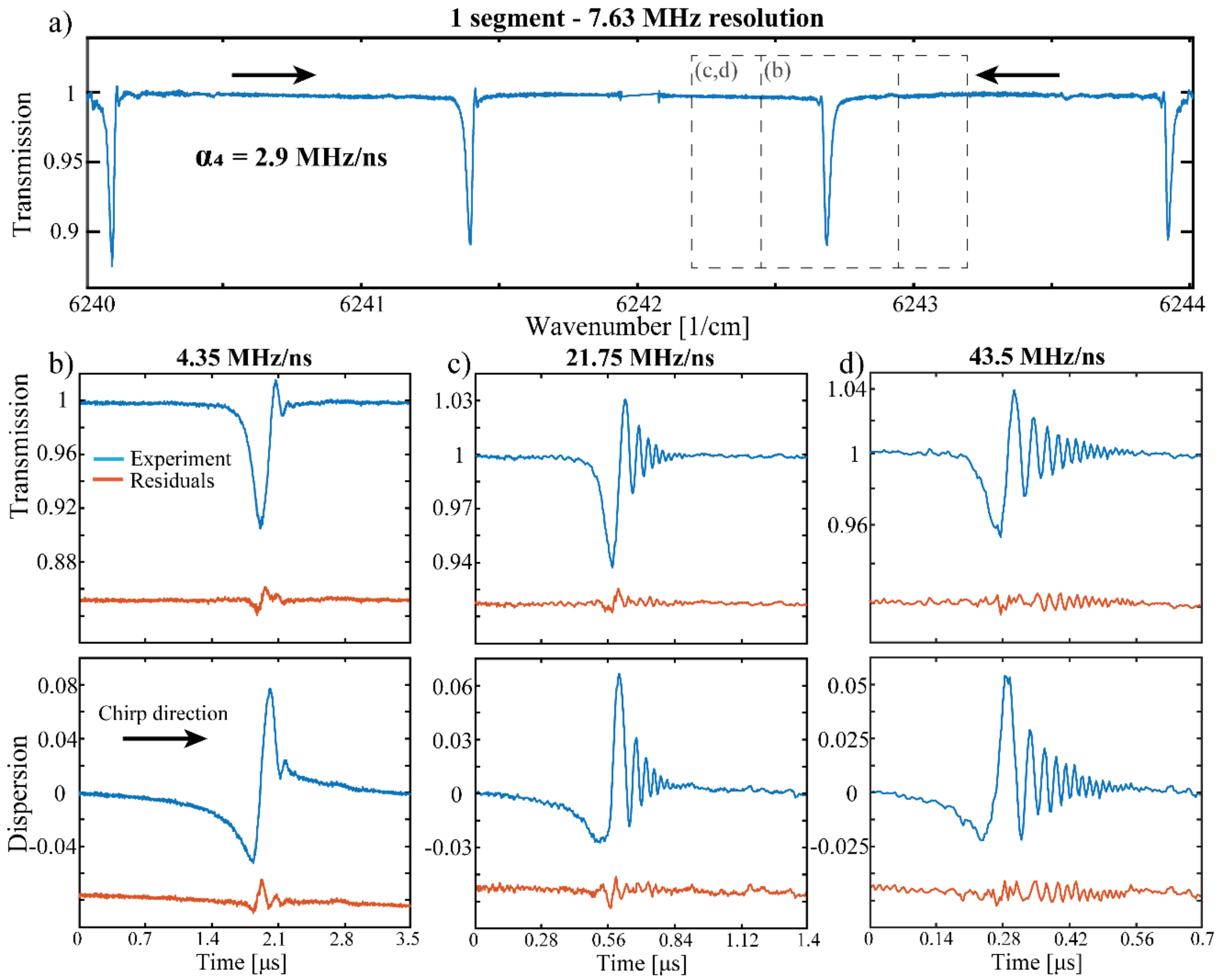

**Figure 5**. The rapid passage spectra of a) four $CO_2$ absorption lines (30013←00001, R16e, R18e, R20e, R22e, left-to-right) at 6.6 kPa (50 torr) (top panel). The observed amplitude (middle panels) and dispersive responses (lower panels) from 3rd-order scans of the R 20e absorption line at 6242.6721 cm$^{-1}$ are shown for increasing scan rates from b) 4.35, c) 21.75 to d) 43.5 MHz/ns. For c) and d), the complex line shape functions are nearly completely transformed to a linear rapid passage response. The residuals (x1) from the Maxwell-Bloch fits are shown below the experimental data in the lower panels.

## 4.4 Acquisition of rapid passage spectra

As the scan rate is increased to greater than 1 MHz/ns, the frequency sweep through resonance leads to asymmetry and post resonance oscillations on the line shape. For the oppositely scanned (+) and (-) sidebands in Fig. 5a, these oscillations appear on opposite sides of the lines as expected. The distortions result from the across-resonance sweep rate that exceeds the collisional relaxation rate for population transfer in a two-level system [62]. The effects of rapid passage have been treated in several studies that use quantum cascade lasers (QCL) [50,51,53,54,63]. We note that the maximum sweep rates used here are more typical of those used in pulsed QCL studies [51,53,54] (linear rapid passage) which are more than 10-fold larger than optimal for population transfer with a cw-QCL [50] (adiabatic rapid passage). However, both of these limits are shown in Fig. 5.

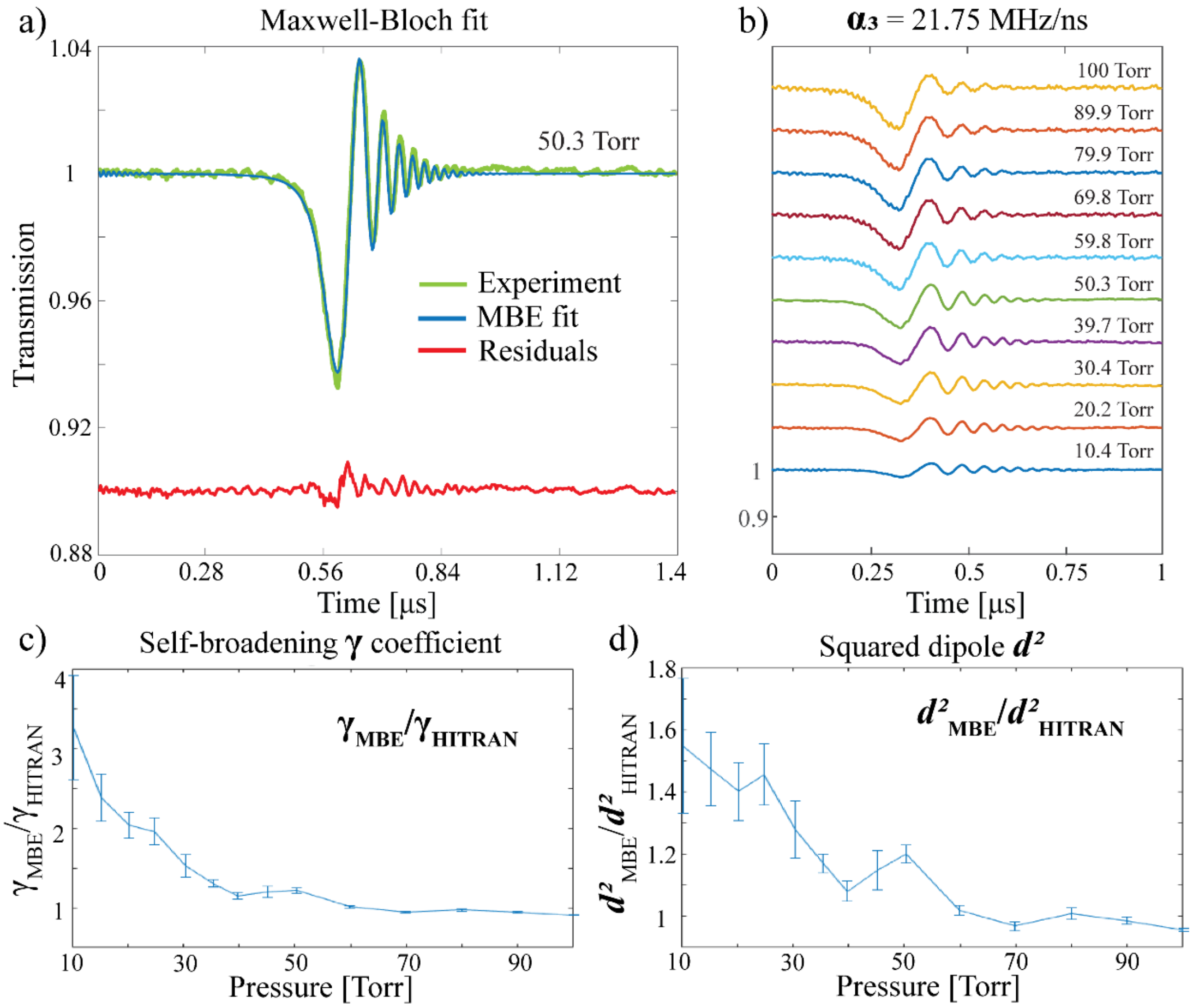

**Figure 6**. a) Shown are the rapid passage signals (green) fit to the MBE model (blue) and the residuals (red) for the $CO_2$ absorption line at 6242.6721 $cm^{-1}$ (R 20e) and 6.6 kPa (50 torr). b) The rapid passage spectra taken at different pressures that increase from bottom to top from 1.3 kPa to 13 kPa (10 torr to 100 torr). The parameters fit to the MBE model which are simple scalars to the HITRAN database parameters that describe the normal line shape in Figs. 3 and 4 for c) the self-broadening coefficient, and d) the squared dipole moment, $d^2$.

Upon decreasing the chirp duration to $\tau_{CP}$ = 20 μs and sweeping from $f_{start}$ = 500 MHz to $f_{stop}$ = 15 GHz, the 3rd- and 4th-order scan rates are $\alpha_3 \approx$ 2.17 MHz/ns and $\alpha_4 \approx$ 2.90 MHz/ns, respectively, where

$$\alpha_m = \frac{m\Delta f_{chirp}}{\tau_{CP}} \qquad (13)$$

In this configuration, the same spectral region is covered, but the resolution has decreased by a factor of 10 from the previous results in Fig. 4c since the chirp duration is decreased by the same amount.

**Table 1. Best fit parameters of the MBE model to the data shown in Fig. 5 (see text for details). Uncertainties represent Type A, k=1 or 1 σ and are reported in the last digit.**

| Parameter | Fig. 5b | | Fig. 5c | | Fig. 5d | |
|---|---|---|---|---|---|---|
| | Transmission | Dispersion | Transmission | Dispersion | Transmission | Dispersion |
| $\gamma/\gamma_{HITRAN}$ | 1.17(4) | 1.091(6) | 1.22(4) | 1.223(8) | 1.27(2) | 1.31(2) |
| $d^2/d^2_{HITRAN}$ | 1.21(3) | 1.194(7) | 1.20(3) | 1.232(7) | 1.22(2) | 1.23(1) |

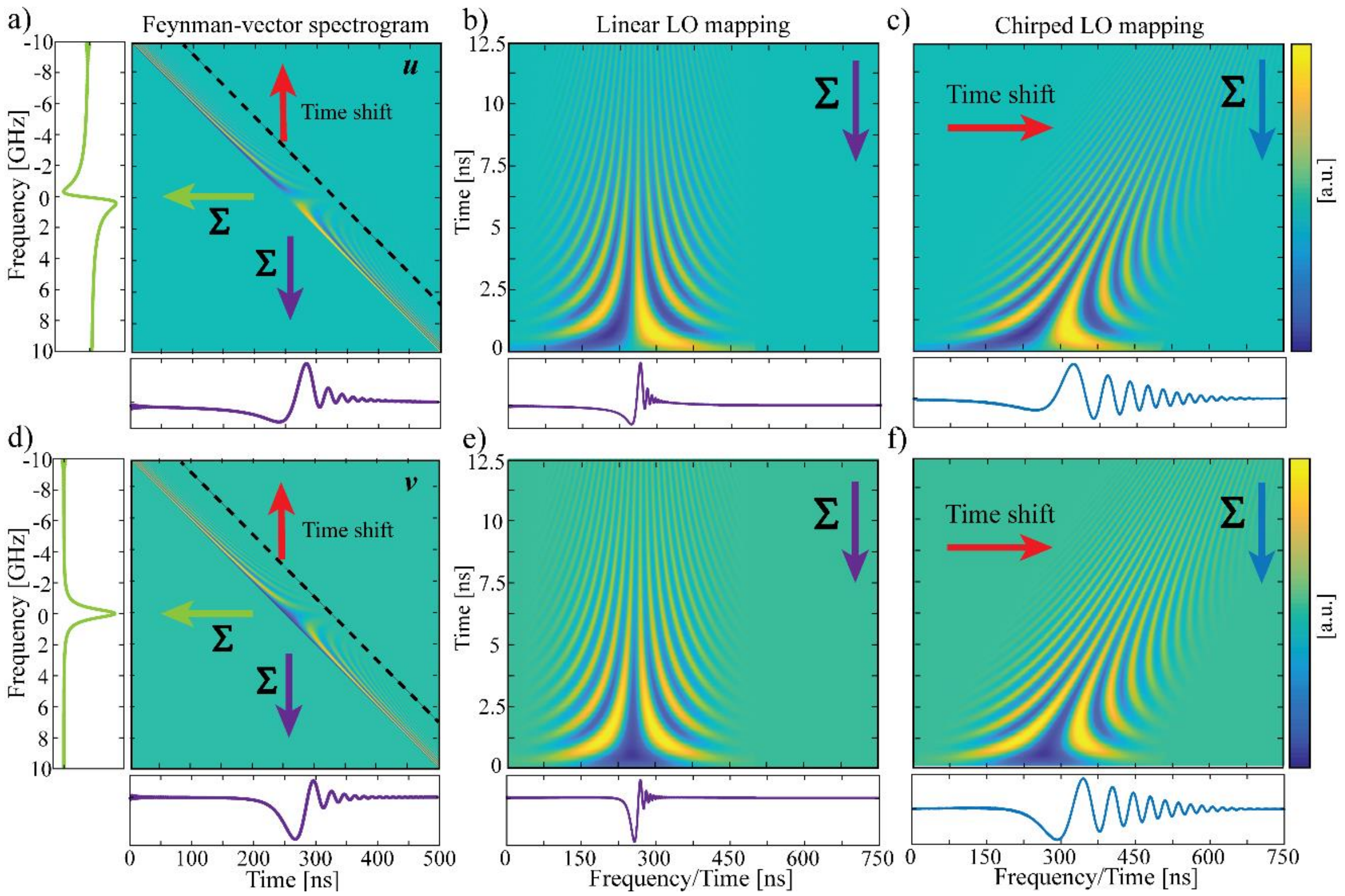

**Figure 7**. The calculated spectral and temporal response of a two-level system when a rapidly chirped pulse is applied. The spectrograms of the Feynman-Bloch vector components of a) *u* (dispersive) and d) *v* (amplitude – see Supplemental Materials for details) show how the system responds at different frequency detunings, Δ, to generate temporal signals that change the observed response of the down converted spectrum. In panels b) and e), a linear mapping of time to frequency is calculated for a fixed LO and temporal responses are shown as vertical sums in the lower sub-panels. In c) and f), the non-linear mapping of the temporal response is calculated for a chirped LO which corresponds to a horizontal time 'shear' applied to the linear map in b) and e). As illustrated by the sums in the lower sub-panels of c) and f), the responses are expanded in time, qualitatively accounting for the chirp rate magnification observed.

As the sweep rate is increased further by decreasing $\tau_{CP}$, the distortions in both the real (middle panel) and imaginary (bottom panel) line shapes become increasingly more pronounced. This is particularly evident for the highest chirp rate of 43.5 MHz/ns in Fig. 5d. By lowering the chirp rate from $\tau_{CP}$ = 10 μs to $\tau_{CP}$ = 1 μs, the effects of rapid passage are seen to completely transform the absorption and phase response.

To account for this behavior, we use the Maxwell-Bloch equations (MBEs) to model the dynamics of a two-level system [64]. The MBEs have recently been used to model dynamics in studies using quantum cascade lasers [50,51,53,54]. Details of the MBE model are given in the Supplemental Materials.

The nonlinear least-squares Levenberg-Marquardt algorithm is used to perform fits of the MBE model to the observed rapid passage data. The two parameters fit are simple scalars to the HITRAN parameters [65] for self-broadening, $\gamma$, and the square of the dipole moment, $d^2$. Data from the 3rd-order scans are used since they have the highest *SNR*. The fits together with the residuals for the $CO_2$ line are shown in Figs. 5b-5d for the three different scan rates of 4.35 MHz/ ns, 21.75 MHz/ns and 43.5 MHz/ns, respectively. A summary of the best fit parameters is given in Table 1.

The dependence of the model parameters on the $CO_2$ pressure is shown in Fig. 6. For the series shown in Fig. 6b, the chirp rate is fixed for pressures from 1.3 kPa to 13 kPa (10 to 100 torr). The results from the MBE fits (see Fig. 6a) of the scalars to the HITRAN self-broadening and squared dipole moment parameters are shown in Figs. 6c and 6d for the different $CO_2$ pressures. These two molecule specific parameters are determined relative to the HITRAN line shape parameters used to model the data in the absence of rapid passage effects (i.e., Figs. 3 and 4). Three sets of data were taken at each pressure and the error bars in Figs. 6c and 6d represent the standard deviations (Type A, k=1 or 1σ) over these sets. While good agreement is found with the HITRAN parameters at the highest pressures, the observed self-broadening coefficients are overestimated by as much as three-fold at the lowest pressures [51]. Similar but smaller deviations are found in comparisons with squared dipole moments. Again, the reasons for discrepancies are under investigation including the degree to which the instrument resolution impacts these results [54].

It is noted that a large discrepancy is found between the actual chirp rate used in the experiment and chirp rate needed in the model to achieve satisfactory fits in Figs. 4 and 5. A scalar applied to the actual chirp rate was estimated across multiple spectra and found to be ≈ 25.3, suggesting that our current experimental arrangement significantly magnifies the effects of rapid passage.

To arrive at a better understanding of this discrepancy, the spectrograms in Figs. 7a and 7d were generated by evolving the Feynman-Bloch vector components in time for a set of fixed detunings, $\Delta$. As the frequency detuning is scanned across the resonance, the field applied is a short pulse ($\tau \approx 100$ ps) with a phase defined by the signal chirp rate. As expected, no response of the system is found before the field is applied and the temporal averages, shown in the left sub-panels in Figs. 7a and 7d, show the steady state frequency response. The time shear required to correct for a chirped LO depends on the amount of bandwidth used in the down converted IF spectrum and the total optical bandwidth scanned. The mapping is done by selecting the time/frequency response indicated by the diagonal dashed line in Figs. 7a and 7d, and the time shift is applied in the direction shown by the red arrows. The spectral response using the chirped LO (Figs. 7c and 7f) effectively increases the chirp rate response of the system relative to a fixed LO (Figs. 7b and 7d), causing the large discrepancy between the chirp rated used in the experiment and chirp rate used in the fits of the MBE model. The difference responses between the fixed LO vs the chirped LO are nicely illustrated by the sums along the detuning axis shown in lower sub-panels of Figs. 7b, 7e and Figs. 7c, 7f, respectively. Although the reason for the magnitude of the scalar is not yet known, it is clearly advantageous to chirp the LO at the expense of comb resolution to better resolve the molecular dynamics of $CO_2$ at these fast chirp rates.

## 5. Conclusion

In this work, a technique to interleave the orders of EOM's using a dual-channel arbitrary waveform generator has been shown to increase the agility of chirped-pulse dual-comb methods. The scheme is implemented by adding a phase slip to the LO leg that enables a unique near-IR-to-RF mapping per order for bandwidth expansion without loss of comb resolution. The expansion to $4^{th}$-order is used to obtain the complex transmission spectrum of four $CO_2$ lines over a 120 GHz bandwidth in the 1.6 μm region. Linear chirped waveforms of variable amplitude are used to improve comb flatness over this spectral region.

Another aspect of chirped pulse waveforms has been exploited to perform molecular dynamics studies of $CO_2$ at high scan rates. While scan rate is expected to scale with the comb order over the expanded bandwidth for a fixed LO frequency, we have further increased the scan rate by more than 1000-fold using a dual chirp technique while maintaining the unique down-conversion of the orders into a 500 MHz RF region. Scan rates from 0.01 MHz/ns to 43.5 MHz/ns have been used to follow the transformation of the $CO_2$ transmission signals from the normal absorption/dispersion line shape to a linear rapid passage signal response [64]. The rapid passage effects have been modeled using the Maxwell-Bloch formalism to obtain parameters related to the relaxation time (Lorentzian width in frequency domain) and the electric dipole transition moment of a $CO_2$ line. Comparisons with the HITRAN parameters show deviations at low pressure which are not yet fully understood. Modifications to the MBE model are currently underway for $CO_2$ and for the more complicated rapid passage effects observed for $CH_4$ in the 1.65 μm region.

It is further noted that the ability to easily modify the linear chirp rate, spectral bandwidth and comb resolution while maintaining high spectral purity to perform molecular dynamics studies across a wide range of time scales and conditions would otherwise be difficult to impossible to perform using conventional single-frequency rapid-scanning techniques [66,50,51,53,54]. Furthermore, while these studies were performed up to 6.6 kPa (50 torr) of $CO_2$ in a short pathlength cell, higher chirp rates are possible by increasing the bandwidth in the IF region [67]. A 5-fold increase in chirp rate may allow investigations at atmospheric pressure for applications to remote sensing.

Efforts are also on-going to improve sensitivity by using dual-drive Mach-Zehnder modulators, to expand coverage through non-linear processes after amplification using EDFAs and to adapt the methods for multiheterodyne IPDA and DIAL studies of GHGs. The enhanced bandwidth and rapid scan rate of dual comb methods can be advantageous for pump-probe molecular dynamics experiments and for quantum coherence studies in cold molecular beams [35,37]. Finally, simple modifications to this method open the way to improve sensitivity using background-free detection methods [21,55].

## Acknowledgements

Support was provided by the NIST Greenhouse Gas Measurements and Climate Research Program managed by James Whetstone. J.B.S. wishes to acknowledge the support from the Summer Undergraduate Research Program (SURF) and J.R.S. wishes to acknowledge the support from the NIST NRC fellowship program. We also wish to acknowledge helpful discussions with Kevin Cossel, Scott Diddams, and Kimberly Briggman.

## Disclosures

The authors do not have any conflicts of interest.

## Data Availability

Data for the results presented in this work are available at the following link.

See supplement document for supporting content.